\documentclass[reprint,aps,prc,twocolumn,superscriptaddress,letterpaper,floatfix,amsmath,amssymb,10pt]{revtex4-2}
\usepackage{graphicx}
\usepackage{color}
\definecolor{navy}{rgb}{0, 0, 0.5}

\usepackage[T1]{fontenc}

\usepackage{xspace}
\usepackage{xr-hyper}
\usepackage{hyperref}
\hypersetup{breaklinks=true,colorlinks=true,linkcolor=blue,citecolor=blue,filecolor=magenta,urlcolor=cyan}

\usepackage[all]{hypcap}

\usepackage{xcolor}
\definecolor{pastelgray}{rgb}{0.81, 0.81, 0.77}
\definecolor{beaublue}{rgb}{0.9, 0.9, 0.93}
\usepackage{orcidlink}

\begin{document}

\title{Quadrupole Strength in Isobaric Triplets}

\author{B. C. Backes\orcidlink{0000-0002-4490-2802}}
\email{betania.backes@york.ac.uk}
\affiliation{School of Physics, Engineering and Technology, University of York, Heslington, York YO10 5DD, United Kingdom}
\author{J. Dobaczewski\orcidlink{0000-0002-4158-3770}}
\affiliation{School of Physics, Engineering and Technology, University of York, Heslington, York YO10 5DD, United Kingdom}
\affiliation{Institute of Theoretical Physics, Faculty of Physics, University of Warsaw, ul. Pasteura 5, PL-02-093 Warsaw, Poland}
\author{D. Muir}
\affiliation{School of Physics, Engineering and Technology, University of York, Heslington, York YO10 5DD, United Kingdom}

\author{W. Nazarewicz\orcidlink{0000-0002-8084-7425}}
\affiliation{Facility for Rare Isotope Beams and Department of Physics and Astronomy, Michigan State University, East Lansing, Michigan 48824, USA}

\author{P.-G. Reinhard\orcidlink{0000-0002-4505-1552}}
\affiliation{Institut f\"ur Theoretische Physik, Universit\"at Erlangen, Erlangen, Germany}

\author{M. A. Bentley\orcidlink{0000-0001-8401-3455}}
\affiliation{School of Physics, Engineering and Technology, University of York, Heslington, York YO10 5DD, United Kingdom}

\author{R. Wadsworth\orcidlink{0000-0002-4187-3102}}
\affiliation{School of Physics, Engineering and Technology, University of York, Heslington, York YO10 5DD, United Kingdom}

\date{\today}
\begin{abstract}
The dependence of the $E2$ matrix elements on isospin projection $T_z$ is linked to the conservation of the isospin symmetry. To study this conjecture, we calculated the  ${B(E2: 2^+ \rightarrow 0^+)}$ rates for the even-even $T=1$ mirror nuclei with 42 ${\leq}$ ${A}$ $\leq$ 98 within nuclear density functional theory, employing the generalized Bohr Hamiltonian, and carrying out angular momentum projection. We demonstrated that collective effects are crucial for describing experimental data near the $N=Z$ line without invoking explicit beyond-Coulomb isospin symmetry-breaking corrections. We also determined the $B(E2\downarrow)$ values for odd-odd $T_z=0$ nuclei $^{70}$Br and $^{78}$Y in  doubly-blocked configurations. We discussed the requirements for accurately describing isobaric analog states and emphasized how current theoretical results should be interpreted within the study of isospin symmetry across isospin triplets.
\end{abstract}

\maketitle

\section{Introduction}

The studies of nuclei in the vicinity of  the  $N = Z$ line offer unique opportunities  to test  nuclear models and the degree of isospin-symmetry breaking in nuclei. In $N \approx Z$ nuclei, the large overlaps of neutron and proton wave functions enhance the $T=0$ part of the neutron-proton ($np$) interaction, a driver of nuclear collectivity. Indeed, some of the most collective low-lying nuclear states are found  along the $N=Z$ line in the $A\sim80$ region~\cite{(Lle20)}. 

A key driver for studies of nuclei around $N=Z$ is the dependence of  nuclear structure effects on  isospin \cite{Lenzi2009,(Sat16f),Bentley2022,Smirnova2023}. In members of an isobaric multiplet, isobaric analog states (IAS) exist in  isobars with different  isospin projections, $T_z=\frac{N-Z}{2}$. The IAS are structurally very similar, with any small differences in wave functions attributed to the effect of isospin-breaking interactions such as those of electromagnetic origin. In particular, isobaric triplets with $T=1$ and $T_z=0,\pm 1$ (isotriplets) have been an active area of  studies of isospin-related effects since a detailed spectroscopy (energies and transition strengths)  is experimentally possible for all three members of the isomultiplet.

This is especially the case where the two $T_z=\pm 1$ isobars are even-even nuclei and the $T_z=0$ isobar is an odd-odd system. This is because, in the odd-odd $N=Z$ isobar, the $T=0$ and $T=1$ states compete at very low excitation energy. Indeed, it is only in the  odd-odd $N=Z$ nuclei with $T=1$ ground states that  the nuclear ground state has $T>|T_z|$. During the recent years, isotriplets  have been the target of experimental studies~\cite{Wimmer21,Giles,Boso,(Zim24),(Zim25)} focusing  on the variation of ``analog'' $B(E2)$ transition strengths with $T_z$ -- i.e. the transitions between the $T=1, J^\pi=2^+$  excited states and the  $T=1, J^\pi=0^+$ ground states. Probing the variation of  $B(E2)$ rates across isotriplets is the focus of this theoretical work, a study that enables the investigation of possible structural variations along the isotriplet. 

\section{General Considerations}\label{sec:general}

In the limit of pure isospin symmetry (i.e., in the absence of isospin-breaking interactions), there are exact predictions for the variation of analog transitions matrix elements with $T_z$ that come directly from the isospin formalism. These rules originate from the fact that nuclear current density can be separated into an isoscalar and isovector part~\cite{Warb}, which allows  the extraction of  isoscalar and isovector components of matrix elements. 
(Strictly speaking, the $E2$ transitions, which are of main interest in this study, depend on the distribution of the nuclear charge density $\rho_{\rm ch}$ rather than the proton density $\rho_p$. As discussed in Ref.~\cite{Reinhard2021a}, $\rho_{\rm ch}$  is obtained by folding  $\rho_p$ with the nucleonic charge distributions. Additional corrections come from the effect of center-of-mass projection. We estimated these corrections to have practically no effect on the quadrupole moments and $B(E2)$ rates, considering the typical experimental uncertainties. Therefore, in the following, we base our discussion on the proton density.)

For a set of isospin-conserving ($T\rightarrow T$) analog transitions within a multiplet, the variation of the proton $E2$ matrix element with $T_z$ can be written as~\cite{Bernstein}
\begin{equation}\label{MpTz}
    M_p(T_z)= \frac{1}{2} \left [ M_0-M_1T_z\right],
\end{equation}
where $M_0$ and $M_1$ are the isoscalar and isovector components, respectively. Thus, in the limit of  isospin symmetry, the proton matrix element should be  linear in $T_z$ along the iso-multiplet. Conversely, any observed deviation from linearity must imply a breakdown of the IAS isospin purity. For isotriplets with $T_z=0,\pm 1$, the measurement of the three $B(E2)$s, which are proportional to $|M_p(T_z)|^2$, is sufficient to test this rule, by looking for deviations from the predicted linear behaviour.

The linearity test requires three high-precision transition-rate measurements, and this limits the availability of high-quality data. 
Below $A=50$, such an analysis has been carried out in Ref.~\cite{Boso} where it was demonstrated that the decomposition (\ref{MpTz}) works fairly well. However, two recent results for heavier nuclei have indicated deviations from linearity. Firstly, a recent analysis of $B(E2)$ rates in
 the A=70 isotriplet ($^{70}$Kr,$^{70}$Br,$^{70}$Se) \cite{Wimmer21}, with the $B(E2)$ for $^{70}$Kr around 60$\%$ larger than that of $^{70}$Se, and the value in the odd-odd $^{70}$Br significantly reduced. They suggested that shape variation across the isotriplet could be a possible explanation. The second instance is the unexpected behavior in the $B(E2)$ measurements in the two members of the $A=78$ isotriplet~\cite{(Lle20)}, $^{78}$Sr ($T_z=1$) and $^{78}$Y ($T_z=0$), where the $B(E2)$ in $^{78}$Sr is around 50$\%$ larger than in the $N=Z$ system $^{78}$Y. Whilst the proton-rich member of the isotriplet, $^{78}$Zr, still remains to be measured, the continuation of a linear trend across the isotriplet would result in highly asymmetrical $B(E2)$s between the even-even mirrors.

Experimental studies of nuclei near the $N=Z$ line cannot be quantitatively interpreted in terms of the isospin-conserving frameworks, as this symmetry is explicitly broken by the electromagnetic interactions. But the opportunity is there:  detailed experimental information can shed light on   the mechanisms of isospin symmetry breaking. The important questions to ask in this context are: (i) Is the theoretical description of the electromagnetic effects near the $N=Z$ line sufficiently precise?  (ii) What  data contain signatures of the isospin-symmetry breaking interactions beyond the Coulomb force?  (iii) What is the role of isospin in forming low-lying states of the isotriplet? 

Let us first comment on point (i).
Even if the electromagnetic interactions  in nuclei  are significantly weaker than the strong force, their detailed impact on nuclear structure is highly non-perturbative \cite{Satula2011,Satula2012}. Within a variational approach, such as the nuclear DFT  employed in this study, the Coulomb force is fully accounted for. Consequently, near the $N=Z$ line, the deformed single-particle energies and wave functions of neutrons and protons will usually differ. 
As a result, the deformation properties of the $T_z=\pm 1$ members of isotriplets, which are essential for the $B(E2)$ transition analysis, may or may not be similar depending on underlying configurations. 

The Coulomb force is predominantly a combination of isoscalar and isovector terms with a weak isotensor component. Therefore, it contributes to and defines the slope of the linear trend (\ref{MpTz}), and also induces weak departures from the linear behavior. 
The Coulomb force is also believed to impact other components of nuclear interaction, such as pairing \cite{Anguiano2001}. For instance, to accommodate the experimental odd-even mass staggering, 
the effective proton and neutron pairing interactions in atomic nuclei  differ
\cite{Bertsch2009}. Since nuclear low-lying collective modes 
are strongly affected by pairing \cite{Belyaev1959,Bohr75,Brack1972}, one expects to see isospin-violating effects in the structure of 2$^+$ states and $E2$ matrix elements.

\section{Structure of IAS in odd-odd nuclei}\label{sec:IAS}

The standard approach to describing paired odd-$A$ nuclei is within the so-called blocking approximation, whereupon a single quasiparticle is created atop an even-even quasiparticle vacuum. 
Within the quasiparticle  pairing theory, such as BCS or HFB,  this corresponds to the occupation probability of one single-particle state becoming equal to one and that of its canonical partner becoming equal to zero, and to the total energy increased by the corresponding 
pair-breaking contribution \cite{(Rin80b)}. 
The corresponding independent-quasiparticle wave function in  a canonical-HFB representation
can be written as:
\begin{equation}\label{oddA}
  |\Psi_{\alpha_0\tau_0}\rangle =a^\dagger_{\alpha_0\tau_0}
  \sideset{}{'}\prod_{\alpha\tau\ne\alpha_0\tau_0} (u_{\alpha\tau} + v_{\alpha\tau}
  a^\dagger_{\alpha\tau} a^\dagger_{\bar\alpha\tau} )|0\rangle,
\end{equation}
where $\alpha_0$ represents space-spin quantum numbers of a blocked state (e.g.\ its Nilsson labels) and $\tau_0$ represents its isospin projection $n$ or $p$,
$u_{\alpha\tau}$ and $v_{\alpha\tau}$ are canonical occupation 
 coefficients, $\bar\alpha$ represents canonical partner state of state $\alpha$, and $|0\rangle$ is the single-particle vacuum. The prime symbol indicates that the product runs over one-half of the states, customarily denoted by
 $\alpha >0$, that is, the partner states $\bar\alpha$ are excluded from the product. Evidently, for $\tau_0=n$ or $p$, state (\ref{oddA}) represents the state of an odd-$N$ or odd-$Z$ nucleus, respectively.

Similarly, one would think that states of odd-odd nuclei can be approximated by blocking one proton
quasiparticle in state $\alpha_{0}p$ and one neutron quasiparticle in  state ${\bar\alpha}_{0}n$, i.e.,
\begin{equation}\label{oddoddA}
  |\Psi_{\alpha_{0}n,{\bar\alpha}_{0}p}\rangle =a^\dagger_{\alpha_{0}p}
  a^\dagger_{{\bar\alpha}_{0}n}
    \sideset{}{'}\prod_{\alpha\tau\ne\alpha_{0}p,{\alpha}_{0}n}(u_{\alpha\tau} + v_{\alpha\tau}
  a^\dagger_{\alpha\tau} a^\dagger_{\bar\alpha\tau} )|0\rangle,
\end{equation}
where the selection of the blocked partner state ${\bar\alpha}_{0}p$, characterized by an opposite projection of angular momentum on the axial symmetry axis, aims to model the $J_z=0$ states in odd-odd nuclei.

The doubly-blocked states (\ref{oddoddA}) can form a useful basis to describe odd-odd nuclei.
Therefore, in the self-consistent theory, the total energy contains neutron and proton pair-breaking energies corrected by contributions from the proton-neutron interaction between the blocked particles.

When approaching the $N=Z$ line, however, things get complicated. 
Consider the hypothetical $T=1,T_z=0$ IAS state in an odd-odd nucleus with $N=Z$  obtained by acting on the ground state of an even-even $T_z=\mp 1$  nucleus with the isospin ladder operator $\hat{T}_\pm$:
\begin{equation}\label{IAS}
|\Psi_{\rm IAS}(T=1,T_z=0)\rangle=\frac{1}{\sqrt{2}}
\hat{T}_\pm|\Psi_{\rm e-e}(T=1,T_z=\mp 1)\rangle.
\end{equation}
To gain insights into the structure of the IAS state in the single-particle (s.p.) basis, we assume that the underlying Hamiltonian is isospin invariant, which implies that the s.p.
proton and neutron wave functions are identical.
By taking the isospin ladder operators in the usual form
\begin{equation}
    \hat{T}_+^\dagger=\hat{T}_-=\sideset{}{'}\sum_\alpha \left(a_{\alpha p}^\dagger a_{\alpha n}+a_{\bar\alpha p}^\dagger a_{\bar\alpha n}\right),
\end{equation}
and approximating $|\Psi_{\rm e-e}(T=1,T_z=\mp 1)\rangle$
by the quasiparticle vacuum
|$\Psi_{\rm vac}^\mp\rangle=    \sideset{}{'}\prod_{\alpha\tau}(u_{\alpha\tau}^\mp + v_{\alpha\tau}^\mp
  a^\dagger_{\alpha\tau} a^\dagger_{\bar\alpha\tau} )|0\rangle$, one obtains \cite{Cwiok1980}
 \begin{eqnarray}\label{PsiIAS}
   |\Psi_{\rm IAS}\rangle\!\!&=&\!\!\sideset{}{'}\sum_\alpha u_{\alpha p}^+v_{\alpha n}^+
   {\cal S}^{T=1}_\alpha 
     \sideset{}{'}\prod_{\alpha\tau\ne\alpha n,\alpha p}(u_{\alpha\tau}^+ + v_{\alpha\tau}^+
  a^\dagger_{\alpha\tau} a^\dagger_{\bar\alpha\tau} )|0\rangle \nonumber \\
                         \!\!&=&\!\!\sideset{}{'}\sum_\alpha u_{\alpha n}^-v_{\alpha p}^-
   {\cal S}^{T=1}_\alpha 
     \sideset{}{'}\prod_{\alpha\tau\ne\alpha n,\alpha p}(u_{\alpha\tau}^- + v_{\alpha\tau}^-
  a^\dagger_{\alpha\tau} a^\dagger_{\bar\alpha\tau} )|0\rangle,\nonumber \\
\end{eqnarray} 
where
\begin{equation}
   {\cal S}^{T=1}_\alpha =  \frac{1}{\sqrt{2}} 
   \left(a^\dagger_{\alpha_p}
  a^\dagger_{\bar\alpha_n} + 
  a^\dagger_{\alpha_n}a^\dagger_{\bar\alpha_p}
   \right)
\end{equation}
represents a time-even proton-neutron pair with $T=1$.
We note that the equality of two forms of $|\Psi_{\rm IAS}\rangle$ given in Eq.~(\ref{PsiIAS}), which reflects the analogous equality in  Eq.~(\ref{IAS}), is based on symmetry of the occupation factors
$v_{\alpha p}^-=v_{\alpha n}^+$ and $v_{\alpha n}^-=v_{\alpha p}^+$ in the $T_z=\pm 1$ mirror nuclei.

By inspecting Eq.~(\ref{PsiIAS}), one concludes that the IAS wave function {\em is not} a product state associated with a doubly blocked even-even vacuum (\ref{oddoddA}) but rather a coherent superposition of the blocked $T=1$ proton-neutron pairs atop the even-even quasiparticle vacua. However, the contributions from individual components are weighted by factors $u_{\alpha p}^+v_{\alpha n}^+=u_{\alpha n}^-v_{\alpha p}^-$, i.e., they are peaked around the corresponding neutron and proton Fermi levels. It is now clear that by taking a symmetrised combination of the double-blocked states (\ref{oddoddA}) at the Fermi levels, we obtain not a complete IAS (\ref{PsiIAS}) but only its dominating component. Note that in the absence of pairing, the sum in (\ref{PsiIAS}) will be reduced to only one term associated with the valence neutron level $\alpha$ occupied in the $T_z=1$ nucleus ($v_{\alpha n}^+$=1) with the proton level $\alpha$ being empty ($u_{\alpha p}^+=1$) (the opposite holds for the $T_z=-1$ system).

A similar interpretation of IAS $T=1$ states in odd-odd $N=Z$  nuclei has been offered by the 
iso-cranking approach
\cite{Satula2001,Glowacz2004} based on the so-called false vacuum
situated mid-way between
the vacua of even-even $T_z=\pm 1$ neighbors.

\section{Theoretical Models}

This section summarizes the theoretical approaches employed in this work. While they are all based on nuclear DFT, they differ in how the low-energy $B(E2)$ rates were extracted from intrinsic self-consistent densities.

\subsection{The Generalized Bohr Hamiltonian\label{GBH}}

To describe quadrupole collective excitations and determine the values of ${B(E2: 2^+ \rightarrow 0^+)}$, we utilized the Generalized Bohr Hamiltonian (GBH)~\cite{row10,Kluepfel2008,pro09}, which encompasses the rotational and vibrational motion associated with quadrupole deformations ($\beta, \gamma$). The general form of the GBH is given by: 
\begin{align}\label{collective}
\hat{H}_{coll}&=\hat{T}_{vib}+\hat{T}_{rot}+\hat{V}\left(\beta,\gamma\right)\;,
\end{align}
\noindent where
\begin{align}
\hat{T}_{\text{vib}}&=-\frac{1}{2\sqrt{wr}}\left\{\frac{1}{\beta^{4}}\left[\partial_{\beta}\left(\beta^{4}\sqrt{\frac{r}{w}}{B_{\gamma\gamma}}\right)\partial_{\beta}\right.\right. \nonumber \\
&\left. -\partial_{\beta}\left(\beta^{3}\sqrt{\frac{r}{w}}{B_{\beta\gamma}}\right)\partial_{\gamma}\right] \nonumber \\
& +\frac{1}{\beta\sin\left(3\gamma\right)}\left[-\partial_{\gamma}\left(\sqrt{\frac{r}{w}}\sin\left(3\gamma\right){B_{\beta\gamma}}\right)\partial_{\beta}\right. \nonumber \\
&\left.\left. +\frac{1}{\beta}\partial_{\gamma}\left(\sqrt{\frac{r}{w}}\sin\left(3\gamma\right){B_{\beta\beta}}\right)\partial_{\gamma}\right]\right\}\;, \label{Tvib}\\
\hat{T}_{\text{rot}}&=\frac{1}{2}\sum^{3}_{k=1}\frac{I^{2}_{k}\left(\Omega\right)}{4B_{k}\left(\beta,\gamma\right)\beta^{2}\sin^{2}\left(\gamma-2\pi k/3\right)},
\label{Trot}
\end{align}
where $\hat{V}\left(\beta,\gamma\right)$ is the potential energy of the nucleus, $B_k\, (k=1,2,3)$ are 3 components of the rotational moment of inertia,
$B_{\beta\beta}, B_{\beta\gamma}, B_{\beta\gamma}$ are components of the quadrupole vibrational ineria tensor,
and $I_k$  denotes the k-component of the angular momentum in the nuclear body-fixed frame. All inertia (mass parameters)   are parametrised in terms of the Bohr deformations $\beta$ and $\gamma$. In Eq.~(\ref{Tvib}), we  used the  shorthand notation:
$w=B_{\beta\beta}B_{\gamma\gamma}-B^{2}_{\beta\gamma}$ and
$r=B_{x}B_{y}B_{z}.$
The 6 mass parameters used in the GBH have been systematically obtained via the cranking
approximation~\cite{gir79,bar11} from constrained HFB calculations for each nucleus under consideration following the procedure of Ref.~\cite{pro15,Muir:2020sft}.
It is worth noting that the rotational and vibrational inertia strongly depend on pairing \cite{Brack1972,pro09}.

The constrained HFB calculations were performed in the basis of 16 major oscillator shells using the HFODD code~\cite{dob09,(Dob21f),(Dob25)}, version (3.08h), at every point in the $\left(\beta,\gamma\right)$ deformation plane. A step size of $0.05$ in the $\beta$ deformation parameter and $6^{\circ}$ increments in the $\gamma$ deformation parameter adequately cover the $\left(\beta,\gamma\right)$ plane to allow reliable energy spectra and the reduced $B\left(E2\right)$ transition probabilities to be obtained. To account for the neglected time-odd mean fields, this work follows the common practice~\cite{(Vre05c),(Giu18a),(Rys22b)} of uniformly rescaling the inertia by a factor of 1.3. We demonstrate the effect of this approach, citing both scaled and unscaled sets of results; however, the full adiabatic theory must ultimately be employed, as in Ref.~\cite{(Sun25)}, which will be addressed in a forthcoming publication. We refer the reader to Ref.~\cite{pro09} for further discussion on the GBH.

\subsection{Quadrupole collectivity for axial states}

We also performed calculations using the computational scheme given by a
representation of wave functions and fields on the real-space grid in
cylindrical coordinates. 
Here, the  {code \sc SkyAx}~\cite{Reinhard2021} was used to obtain the collective
potential energy surfaces (PES) and collective inertia along the axially-symmetric quadrupole deformation path.
Together with an interpolation into the triaxial $\gamma$ direction, this provides a microscopic determination of the Bohr Hamiltonian (\ref{collective}-\ref{Trot}) in axial symmetry and
allows for the computation of quadrupole correlation corrections to
the ground state and low-lying excitations~\cite{Kluepfel2008,Erler2011}. 
Approximate angular momentum projection is then performed within the
topologically adapted Gaussian Overlap Approximation (topGOA)~\cite{Reinhard1978,Goz85a,Hagino2002,Erler2011}.

The starting point for topGOA
is the raw PES ${\cal V}(\beta)$ which is the energy of a mean-field state $|\Phi_\beta\rangle$ constrained at 
quadrupole deformation $\beta$.  The rotational moment of inertia
$\Theta_\mathrm{rot} \propto B_1$ and the quadrupole collective mass ${\cal
  M}_\beta \propto B_{\beta\beta}$ along the quadrupole deformation path are computed within the
self-consistent cranking  approximation \cite{Reinhard1987}.

A mean-field state $|\Phi_\beta\rangle$ is not an eigenstate of
$\beta$ but embodies some spurious quadrupole fluctuations. The effect
of these fluctuations on the energy is called zero-point energy (ZPE)
which has to be subtracted from the raw 
PES ${\cal V}(\beta)$ before
computing the true collective fluctuations. 
The ZPE-corrected PES can be written as:
\begin{equation}\label{eq:ZPEcor}
  {V}(\beta)
  =
  {\cal V}(\beta)
  -
  \left(E_{{\rm ZPE},\beta}+E_{\rm ZPE,rot}\right),
\end{equation}
where
\begin{eqnarray}
  E_{\rm ZPE,rot}
  &=&
  \frac{\langle{I}_x^2+{I}_y^2\rangle}{2\Theta_\mathrm{rot}}
  g(\langle{I}^2_x\rangle/2)
  \;,
\label{eq:ZPErot}
\\
  E_{{\rm ZPE},\beta}
  &=&
  \frac{\langle\stackrel{\leftarrow}{\partial_{\beta}}
                \stackrel{\rightarrow}{\partial_{\beta}}\rangle}
       {{\cal M}_\beta}
   \left[3-g(\langle{I}^2_x\rangle/2)\right],
\\
  g(\eta)
  &=&
  \frac{\int_0^1 dx\, \eta(x^2-1)e^{\eta(x^2-1)}}
       {\int_0^1 dx\, e^{\eta(x^2-1)}}
  \;.
\end{eqnarray}
 The important ingredient is the  factor $g(\eta)$ which determines the relative contributions of the vibrational ZPE correction $E_{\mathrm{ZPE},\beta}$
and the rotational ZPE correction $E_\mathrm{ZPE,rot}$. It has been derived
within the topGOA \cite{Reinhard1978,Hagino2002}. Tests for simple
nuclei and schematic models have shown that $E_\mathrm{ZPE,rot}$ is a
good approximation to angular momentum projection for $J=0$ for
medium-heavy and heavy nuclei.

The numerical representation in terms of oscillator functions as in
{\sc HFODD} or on a coordinate-space grid as in {\sc SkyAx} makes little
difference if the precision of the calculations is high. There is, however, a basic difference in the handling
of pairing space in both codes. The solver {\sc HFODD} uses a high pairing cutoff
(about 60 MeV) in the
space of quasi-particle energies. The solver {\sc SkyAx} deals with a smaller
cutoff, 5--15 MeV above Fermi energy, and the truncation  is made in the space of natural
orbitals, see, e.g., \cite{Reinhard2021,Kri90a}.

\subsection{Angular momentum projected DFT}\label{AMP}
As an alternative way to determine the values of ${B(E2: 2^+ \rightarrow 0^+)}$, we utilized the code {\sc HFODD}, version (3.28o), \cite{dob09,(Dob21f),(Dob25)} to perform systematic nuclear DFT calculations for all even-even $T=1$ mirror nuclei with 42 $\leq$ $A$ $\leq$ 98 with axial symmetry enforced, followed by the exact angular momentum projection (AMP)~\cite{(She21)}. For these calculations, we increased the proton and neutron pairing strengths by 20\% relative to the original UNEDF1 parametrization to account for the effects of the Lipkin-Nogami approximate particle number projection \cite{Sheikh_2021}.

To this end, we performed a series of HFB calculations constraining the quadrupole moment ${Q_{20}}$ on a grid with a step size of 0.5\,b. After identifying local energy minima on the grid, we carried out unconstrained calculations to find prolate, oblate, or spherical HFB solutions. We then used AMP to project each solution onto good angular momentum $I$ and determined the ${B(E2)}$ rates from the initial and final projected states.
The minima can be characterized using either the dimensionless Bohr deformation parameter $\beta$ or the proton intrinsic quadrupole moments $Q_{20}(p)$, which are conventionally linked as follows,
\begin{equation}\label{beta2}
    \beta=4\pi \frac{\langle r^2 Y_{20} \rangle_p}{3ZR^2} = \frac{\sqrt{5\pi}}{3ZR^2} Q_{20}(p),
\end{equation}
where $ R = 1.2 A^{1/2}$\,fm.

We paid particular attention to the isotriplets with ${A=70}$ and ${A=78}$. To calculate odd-odd ${N=Z}$ nuclei, we first constrained the particle numbers without blocking to obtain the axial false quasiparticle vacua. Then, we broke the time-reversal, signature, and symplex symmetries and used the ${z}$-component of the isoscalar angular frequency vector of ${\hbar\omega_z=0.001}$\,MeV to quantize the single-particle angular momenta along the axial symmetry axis.

As previously discussed, the IAS is a coherent superposition of blocked $T=1$ proton-neutron pairs, with contributions from different states peaking around the Fermi surface. To identify the states at the Fermi surface, we examined the ground states of the closest isotopes (isotones) to the odd-odd nuclei. We connected them to the Nilsson labels of the single-particle neutron (proton) states to be blocked. For neutrons (protons), we then selected states with positive (negative) $z$-projections ${\Omega_z}$ of angular momentum, ensuring that the total angular momentum projections summed to zero.

We note that the doubly blocked states obtained in this way break the signature symmetry, allowing both even and odd angular momenta to be projected from the intrinsic states. However, since the signature operator simply exchanges the blocked neutron and proton quasiparticles, the even and odd angular momenta form the $T=1$ and $T=0$ rotational bands, respectively.
Finally, the doubly blocked HFB solutions were converged and projected onto good $J=0$ and $J=2$ angular momenta. This allowed for the determination of the ${B(E2: 2^+ \rightarrow 0^+)}$ values for individual two-quasiparticle (2-qp) states \eqref{oddoddA} of odd-odd $N=Z$ nuclei~\cite{(Lle20)}.

\section{Results and discussion}

This section discusses our findings. We first present the results for even-even $T=1$ mirrors within the mass range $42 \leq A \leq 98$. Next, we discuss the $A=70$ and $A=78$ triplets.

\begin{figure}[htb!]
    \includegraphics[width=1.0\linewidth]{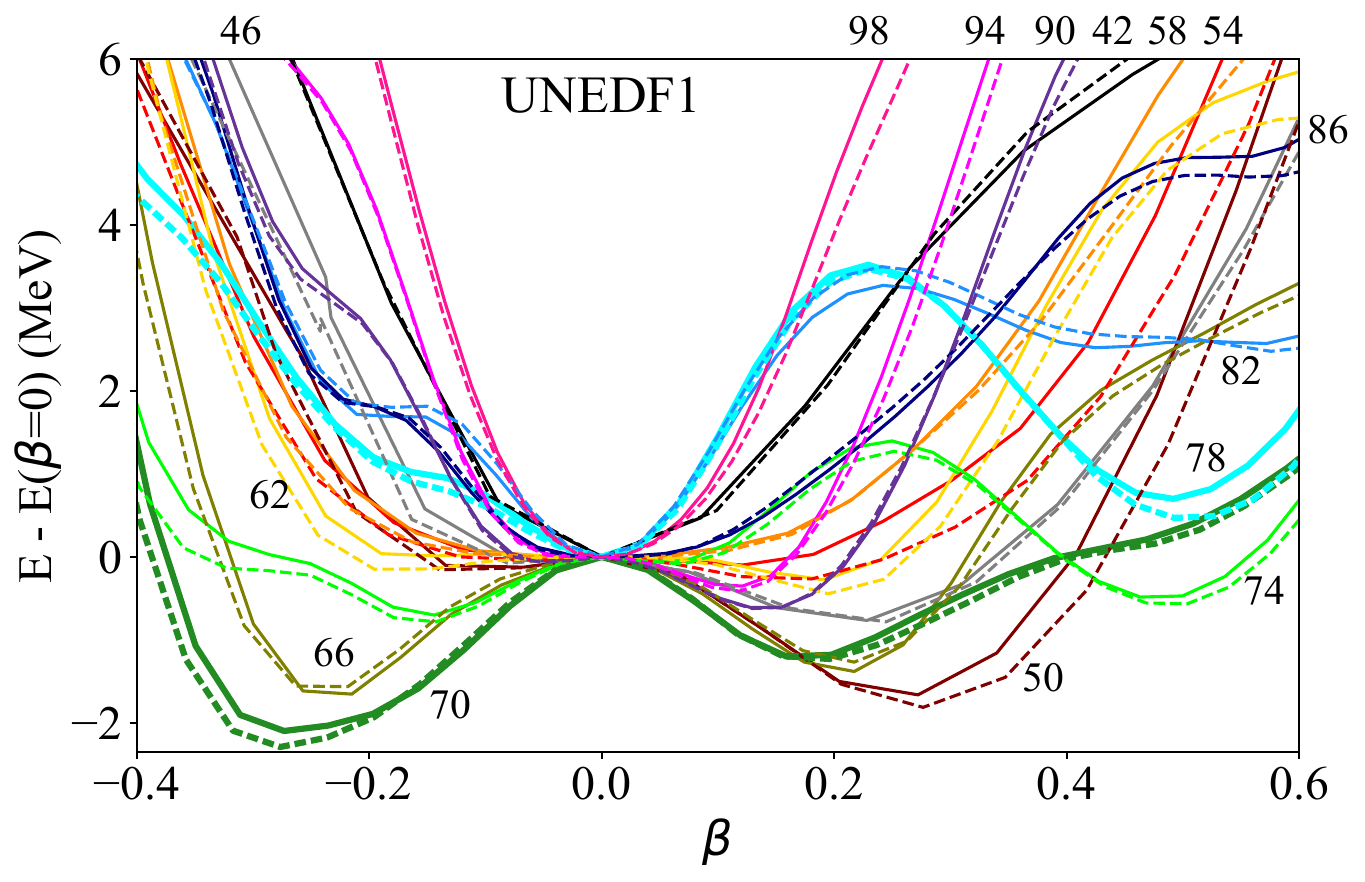}
    \caption{HFB potential energy curves, obtained with UNEDF1, projected on $I=0$  as functions of quadrupole deformation $\beta$ (\ref{beta2}) for the even-even $T=1$ mirror nuclei (solid lines: $T_z=1$; 
    dashed lines: $T_z=-1$) considered in this study. Mass numbers are marked. The energies are normalized to the energy at the spherical shape. The thicker lines show the results for $A=70$ and $A=78$.  
 }
    \label{fig:pes-AMP}
\end{figure}

\subsection{Global analysis of $T=1$ mirror nuclei}

Medium-mass nuclei with $N\approx Z$ are known to exhibit strong coexistence effects with spherical, prolate, and oblate configurations competing energetically at low energies \cite{Nazarewicz1985,ZirconiumRegion1988}. 
Figure \ref{fig:pes-AMP} shows the potential energy curves for even-even mirror nuclei with $42 \le A \le 98$, projected onto angular momentum $I=0$, obtained using the UNEDF1 EDF. According to the calculations, many nuclei are predicted to have shape-coexisting minima. The largest oblate deformations are expected for the $A=66$ and $A=70$ systems, while the largest prolate deformations are calculated for the $A=74$ and $A=78$ nuclei.

The potential energies for $T_z=\pm 1$ mirror nuclei are generally very similar when normalized to the spherical energy. Small deviations, due to the Coulomb interaction, are observed at large deformations, where the energy of the $T_z=-1$ (proton-rich) system falls slightly below that
of the $T_z=1$ partner, which has a smaller atomic number and Coulomb energy. This small shift in total energy results in slightly larger values of $|\beta|$ in $T_z=-1$ mirror partners. Despite the small shifts, shape changes between mirror pairs are unlikely, given the overall similarity between the potential energy curves. In particular, the oblate and prolate minima are separated by $\approx1$ MeV in the $A=70$ mirror pair, which contradicts the suggestion of \cite{Wimmer21} that $^{70}$Kr should be prolate and $^{70}$Se oblate.
We estimated the ${B(E2)}$ transition probabilities based on the predicted energy minima.

\begin{figure}[hbt!]
\includegraphics[width=1.0\linewidth]{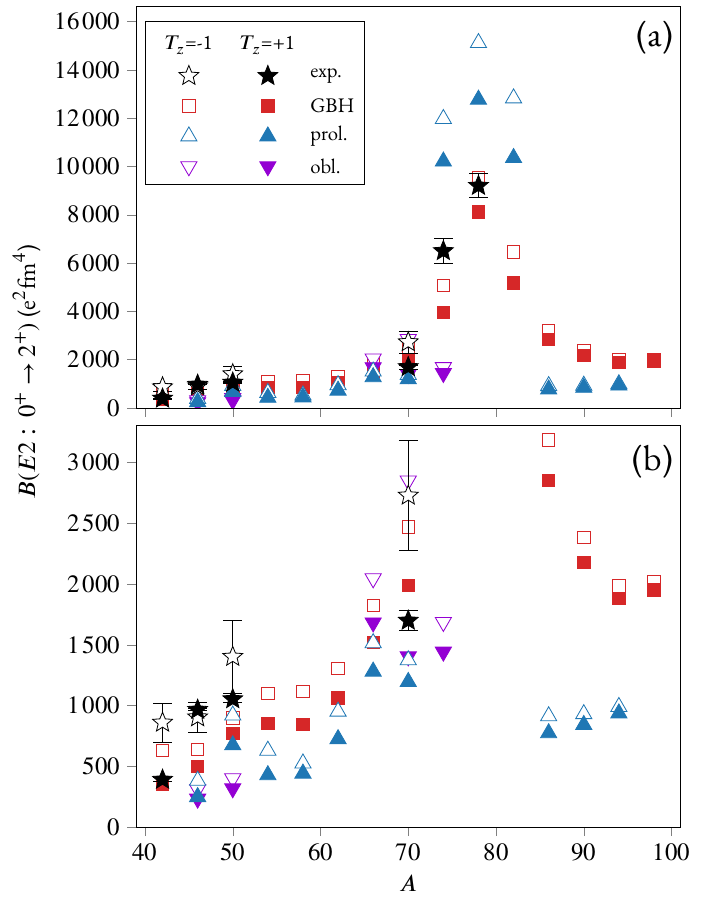}
\caption{\label{fig_BE2-red} ${B(E2)}$ values for the even-even ${T=1}$ mirror nuclei with ${42\leq A \leq 98}$, compared to the experimental data \cite{PRITYCHENKO20161,PhysRevLett.117.062501,Boso,PhysRevC.99.044317,PhysRevC.104.029901,Wimmer21,(Lle20),MORSE2018198}, in (a) full scale and (b) reduced scale. Filled symbols indicate the $T_z=1$ nuclei, while open symbols represent the mirror partners with $T_z=-1$. Experimental values are shown as stars with error bars. The GBH UNEDF1 predictions are marked with squares. The AMP UNEDF1 predictions at prolate and oblate shapes of $A \le 84$ nuclei are marked with up-pointing triangles and down-pointing triangles, respectively.}
\end{figure}

\begin{figure}[hbt!]
\includegraphics[width=1.0\linewidth]{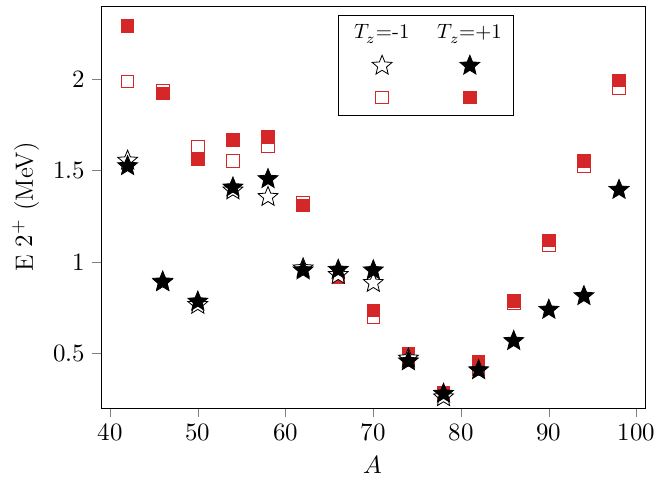}
\caption{\label{fig_E2-red} The energies of the lowest $2^+$ states in even-even $T=1$ mirrow partners. The GBH UNEDF1 predictions (squares) are compared to the experimental data (stars) \cite{ENSDF2,WIMMER2023138249,(Zim25)}.  $T_z=+1$ nuclei are marked with filled symbols and $T_z=-1$ nuclei are marked with open symbols.}
\end{figure}

Figure \ref{fig_BE2-red} presents the results obtained for the ${B(E2)}$ using the GBH and AMP models. Enhanced collectivity is observed in this region of the nuclear chart, and indeed, we see that the GBH calculations, which account for shape mixing, yield better results and demonstrate overall satisfactory agreement with the experimental data across the entire region of interest. For AMP, which does not incorporate shape mixing, we display our results separately for each minima observed in Figure \ref{fig:pes-AMP}. As the AMP calculations fall short in comparison to the GBH and are unable to describe the ${B(E2)}$ in nearly spherical nuclei around shell closures, only results for deformed minima are presented. For nuclei with $74\le A\le 84$, the HFB calculations predict two minima: one nearly spherical and one highly deformed. Experimental data indicate enhanced collectivity in this region and, therefore, significant deformations. However, considering only the prolate-deformed minima, AMP yields ${B(E2)}$ values well in excess of experimental data. This indicates that the spherical-deformed shape mixing may play an important role and that the  GBH approach---which accounts for the distribution of shapes --provides a better description.

In all cases, the  difference 
\begin{equation}
    \Delta {B(E2)}\equiv B(E2,T_Z=-1)
    - B(E2,T_Z=+1)
\end{equation}
is positive. This is to be expected, as the mirror nuclei have very similar equilibrium deformations and the $T_z=-1$  nuclei have larger electric charges. By comparing the results of GBH and AMP, one observes that $\Delta{B(E2)}$ frequently varies between the two models. For heavier systems around $A=82$, this is due to the effects of shape coexistence, as discussed above. For the lighter mirror nuclei, 
such as $A=62$, this is attributed to closely lying weakly deformed prolate and oblate minima.

To complete the overview of GBH results, Fig.~\ref{fig_E2-red} presents predictions for the energy of the first $2^+$ states compared to experimental data. The agreement between experiment and theory is very good for $54\le A\le 86$. Theory overestimates $E(2^+_1)$ values in the remaining cases.  This aligns with the predicted $B(E2)$ values for the $A<54$ mirror nuclei, suggesting reduced quadrupole collectivity compared to the experiment.
The energies of $E(2^+_1)$ states in the mirror partners are very close, with the $T_z=-1$ $E(2^+_1)$ energy generally being slightly lower.
The largest deviations from the isospin symmetry are observed for $A=58$ and $A=70$.

\subsection{The $A=70$ and $A=78$ isobaric triplets}

As we aim to investigate the variation of $B(E2)$ across isotriplets, we will specifically focus on the $A=70$ and $A=78$ triplets, where deviations from the linearity test of Eq.~(\ref{MpTz}) are observed or anticipated.

\begin{figure}[htb!]
    \includegraphics[width=1.0\linewidth]{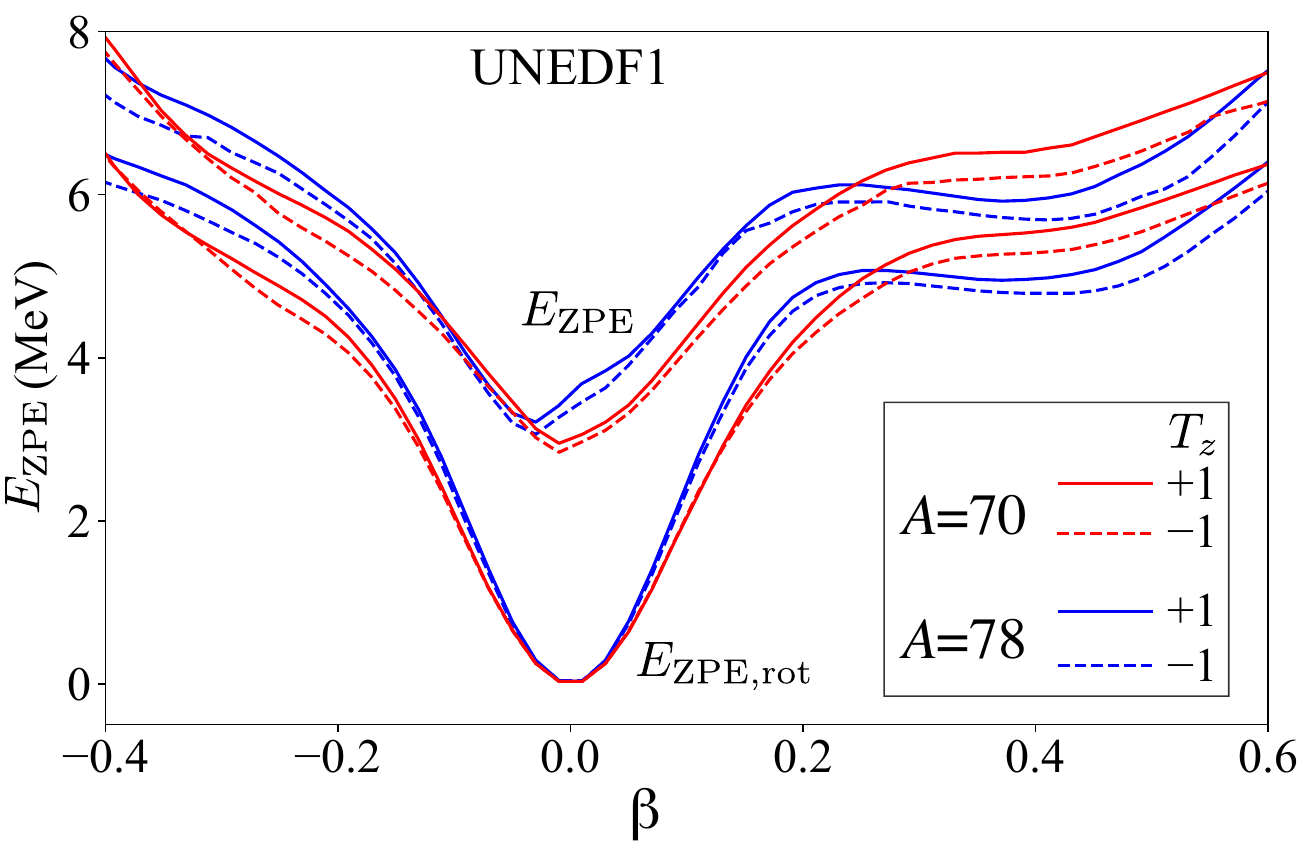}
    \caption{The zero-point energy corrections $E_{\mathrm{ZPE}}$  and $E_{\mathrm{ZPE},rot}$ for the $A=70$ and $A=78$ isobaric triplets.
 }
    \label{fig:pg}
\end{figure}

As shown in Fig.~\ref{fig_BE2-red}, $B(E2)$ values can be influenced by collective effects that extend beyond the single-reference mean-field picture. To illustrate this point, Fig.~\ref{fig:pg}
shows the ZPE (\ref{eq:ZPEcor}) for the $A=70$ and $A=78$ isobaric triplets. The rotational correction $E_{\mathrm{ZPE,rot}}$ is zero at the spherical shape,
increases rapidly at moderate deformations, and then levels off. The difference of $E_{\mathrm{ZPE,rot}}$ between
spherical and deformed minima exceeds 4\,MeV; thus, this represents a significant correction to the total energy. This correction is automatically included in the AMP potential energy curves of Fig.~\ref{fig:pes-AMP}.
As discussed in
Refs.~\cite{Reinhard1987,Reinhard1999}, the rotational ZPE should
be supplemented by the vibrational counterpart $E_{\mathrm{ZPE},\beta}$. This quantity peaks around the spherical shape and reaches a value of about 1\,MeV at large deformations. The resulting total ZPE  $E_{\mathrm{ZPE}}$  increases with deformation and lowers the deformed configurations by 2-3\,MeV compared to the uncorrected calculations. The difference in $E_{\mathrm{ZPE}}$ between the mirror partners is small and steadily increases with deformation.
 We note that the zero-point correction to the quadrupole moment
 is zero as this quantity is used to generate the constrained collective trajectory~\cite{Reinhard1987}.

\begin{figure}[htb!]
\includegraphics[width=\linewidth]{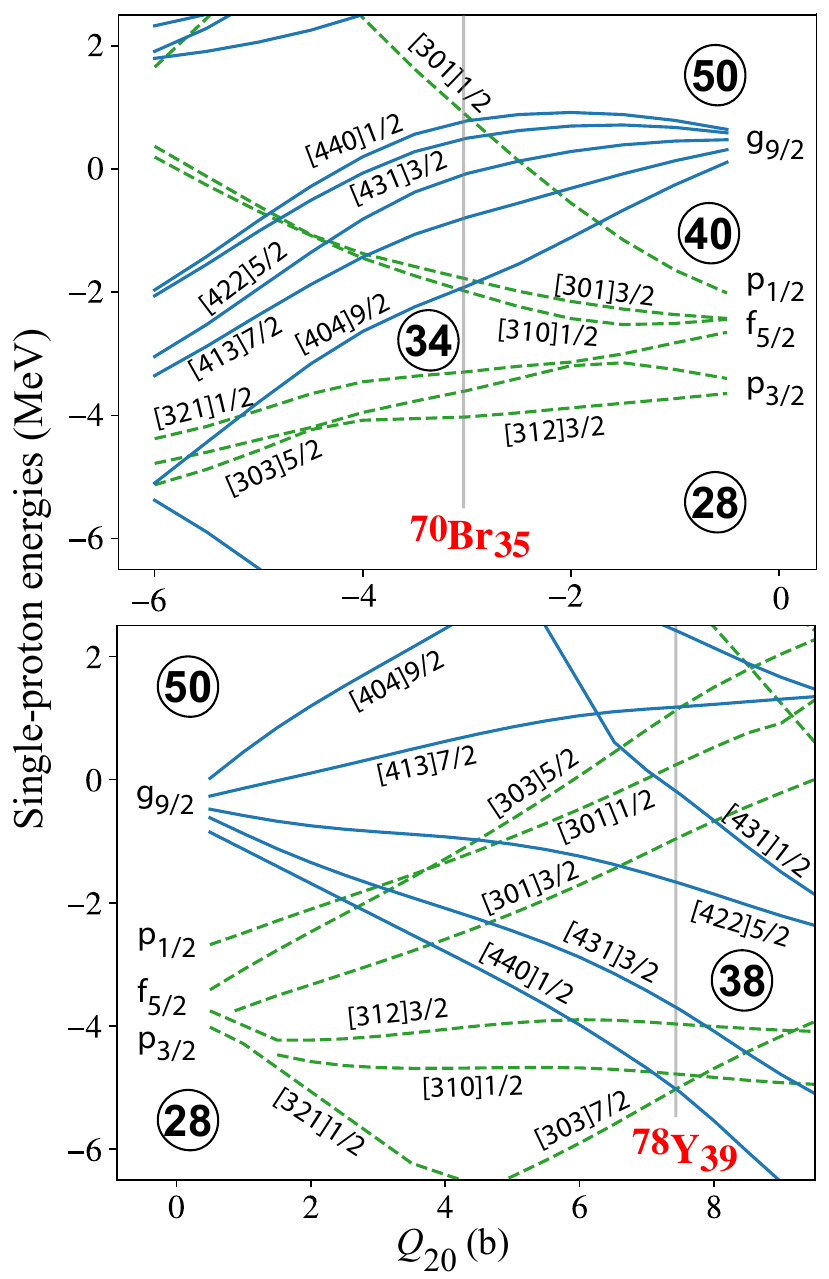}
\caption{\label{fig-Nilsson} The Nilsson proton diagrams for $^{70}$Br (top) and  $^{78}$Y (bottom) obtained using the UNEDF1 EDF and plotted as a function of the total intrinsic quadrupole moment $Q_{20}$.
(The neutron diagrams are quite similar; therefore, they are not shown.) 
Although the wave functions are strongly mixed in many cases, the asymptotic quantum numbers $[Nn_x\Lambda]\Omega$ are provided as in Ref.~\cite{Nazarewicz1985} to simplify the identification of individual levels.
The vertical lines mark the false-vacuum equilibrium quadrupole moments of $A=70$  ($Q_{20}^{\text{min}}\approx-3.2$\,b) and $A=78$  ($Q_{20}^{\text{min}}\approx7.5$\,b) isotriplets.
}
\end{figure}

\begin{figure}[htb!]
\includegraphics[width=1.0\linewidth]{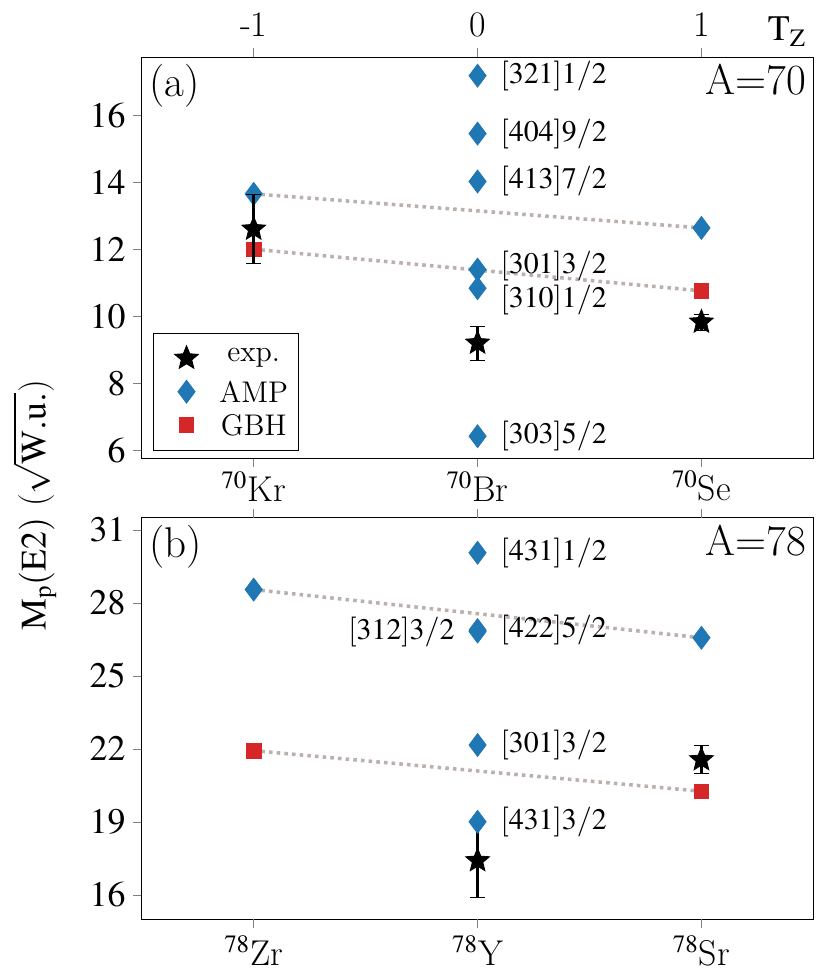} 
\caption{\label{fig_triplets} The values of ${M_p(E2)}$ computed with HFB+AMP using the UNEDF1 EDF for (a) ${A=70}$   and (b)  ${A=78}$ isotriplets, compared with experimental data \cite{Wimmer21}. Different values for the $T_Z=0$ nuclei correspond to various 2-qp states near the Fermi level.}
\end{figure}

 For the analysis of $M_p(T_z)$ (\ref{MpTz}), it is imperative to address the physics of the odd-odd $N=Z$ ($T_z=0$) nuclei. As discussed earlier, the collective degrees of freedom play a crucial role in accurately describing quadrupole strength near the $N=Z$ line, making the GBH formalism the preferred method for describing nuclei in this region. However, the extension of the GBH model to odd-odd nuclei 
that involve decoupled proton-neutron pairs still needs to be developed. We conducted double-blocked calculations using the AMP procedure to gain insights into the $B(E2)$ strengths of the $T_z=0$ isotriplet partners. This method notably falls short in describing spherical nuclei but is, in principle, well-suited for describing deformed nuclei such as $^{70}$Br and $^{78}$Y.

The IAS in the odd-odd mirror nucleus is represented by a superposition of the blocked $T = 1$ proton-neutron pair excitations; see Sec.~\ref{sec:general} and Eq.~(\ref{PsiIAS}).
To analyze this problem, we first computed the $B(E2)$ matrix elements for several 2-qp states near the Fermi level in the oblate configuration of $^{70}$Br and the prolate configuration of $^{78}$Y. 
Figure~\ref{fig-Nilsson} shows the corresponding Nilsson diagrams. It is evident that the main contributions to the IAS wave function  (\ref{PsiIAS}) are expected to arise from [310]1/2, [301]3/2, and [404]9/2 states in $^{70}$Br, as well as from [422]5/2, [301]3/2, and [431]1/2 states in $^{78}$Y.

Figure~\ref{fig_triplets} shows the values of ${M_p}$ for the ${A=70}$ and ${A=78}$ isotriplets considering the same shape for the even-even mirrors. As discussed earlier, the notable difference between the BH and AMP results for the even-even mirror nuclei indicates shape mixing. 
As shown in Fig.~\ref{fig:pes-AMP}, the $A=70$ nuclei are expected to be oblate, while a coexistence of closely lying spherical and prolate shapes is anticipated for $A=78$.
The ${M_p}$ values for $^{70}$Kr, $^{70}$Se, and $^{78}$Sr align with these expectations.The ${M_p}$ values in the 2-qp states of the $T_z=0$ isotriplet members show a significant spread. This is due to the different intrinsic quadrupole moments of associated Nilsson levels, which give rise to different quadrupole polarizations. We also highlight in Fig.~\ref{fig_triplets} the linear trend set by the two $T=1,T_z=\pm1$ nuclei. The different quadrupole polarizations cause the spread of predictions for the odd-odd $N=Z$ nucleus relative to the linear trend.
As none of the 2-qp states uniquely corresponds to the IAS, as discussed in section \ref{sec:IAS}, deviations from the linearity rule (\ref{MpTz}) are expected.

The known or extrapolated ground-state spins and parities of the odd neighbors of $^{70}$Br and $^{78}$Y are $5/2^-$ and $5/2^+$, respectively. This appears to be adequately supported by the Nilsson diagrams in Fig.~\ref{fig-Nilsson}, where the Nilsson levels [303]5/2  and [422]5/2  are located near the Fermi levels of  $N$ or $Z=35$ and $N$ or $Z=39$, respectively. However, for these two Nilsson levels, the departures from linearity shown in Fig.~\ref{fig_triplets} either overestimate or seem to underestimate those visible in the data.

As Eq.~(\ref{PsiIAS}) suggested, a multi-reference mixing of several doubly blocked states should be performed to test whether the IAS hypothesis better aligns with the data. However, our attempts to carry out such a calculation have failed because the resulting energy mixing matrix elements turned out to be strongly perturbed by the self-interaction terms and singularities~\cite{(She21)} of the UNEDF1 functional.
The search for well-adjusted, spectroscopic-quality, and self-interaction-free functionals is actively pursued~\cite{(Sad13),(Ben14b),(Rai14a)}. However, the existing infrastructure is still insufficient to address the $B(E2)$ rates in the IAS of odd-odd nuclei, and the detailed assessment of potential deviations from (\ref{MpTz}) must wait.

\section{Conclusions}

Our theoretical work shows that nuclear DFT describes the reduced transition probabilities of $T=1$ even-even mirrors without explicit beyond-Coulomb isospin symmetry breaking. However, agreement with data is only achieved when accounting for the enhanced collectivity close to the $N=Z$ line by employing a suitable model---in our case, the Generalized Bohr Hamiltonian. Although simple angular momentum projection can reproduce data when the potential energy surfaces have well-defined deformed minima, it struggles with shape coexistence and small deformations close to shell closures.

The DFT results come with a caveat. Although no explicit beyond-Coulomb symmetry breaking exists, collective effects depend on the effective pairing strengths, which differ for protons and neutrons due to the asymmetry between corresponding shell structures across the nuclear chart. Thus, in DFT, differences will always exist between mirror nuclei aside from different Coulomb repulsion.

We discussed which state of the odd-odd nucleus is most suitable for isospin physics studies of isobaric triplets. The true IAS consists of a superposition of doubly blocked $T=1$ proton-neutron pairs atop the even-even quasiparticle vacuum. Our numerical results indicate that double quasiparticle blocking close to the Fermi surface, without mixing, does not lead to the ``linearity rule,'' and the corresponding $B(E2)$ values vary significantly depending on the deformations and slopes of the Nilsson orbitals.
Regarding the linearity rule, we want to emphasize what follows.
\begin{itemize}
    \item While shell model calculations, commonly used to interpret $B(E2)$ data in isobaric triplets,  produce results within the exact isospin symmetry regime, they lack a non-perturbative description of the Coulomb and shape polarizations, which are integral parts of the nuclear DFT;
    \item No shape change is expected between $T=1$ even-even mirrors with $42\leq A \leq 98$, but shape and configuration mixing is a crucial aspect that future DFT implementations need to address;
    \item Analyzing isospin symmetry-breaking from the linear trend set by the $T_z=0,+1$ members of the triplet can be misleading;
    \item Currently, no DFT studies accurately describe the $N=Z$, $T=1$ IAS. The existing infrastructure is not sufficient to capture the complex physics of odd-odd nuclei. Specifically, there is a need for multi-reference calculations using self-interaction-free functionals.
\end{itemize}

Although not addressed in this study, future research on odd-odd $N \sim Z$ nuclei should consider that the neutron-proton overlaps are also anticipated to promote the prevalence of $np$-pairing correlations. The competition between isovector $T=1$ and isoscalar $T=0$ $np$-pairing modes remains a key question in  nuclear structure  physics, and presents nuclear theory with many challenges, see, e.g., Refs.~\cite{frauendorf2014overview,Romero,Engel1997,martinez,SATULA19971} and references therein.

 Within the (quasi)local density approximation to nuclear DFT with pairing, principles of the approach have been laid down in Ref.~\cite{(Per04c)} but have never been implemented or tested so far. The principal difficulty here, which has not yet been widely realized in the previous studies of the $np$-pairing, is that such pairing implies the proton-neutron mixing~\cite{(Sat13d),(She14)} in the particle-hole channel. Indeed, for any quasiparticle vacuum, the density matrix $\rho$ and the pairing tensor $\kappa$ are linked to one another by the expression $\rho-\rho^2=\kappa\kappa^+$~\cite{(Rin80b)}. Therefore, the presence of the three types of pairing, $\kappa_{nn}$, $\kappa_{pp}$, and $\kappa_{np}$, implies, in general, a non-zero proton-neutron mixing $\rho_{np}$ in the density matrix. For nuclear matter, this aspect will be addressed in a forthcoming publication~\cite{(Bac25)}.
\bigskip


\begin{acknowledgments}
This work was partially supported by the STFC Grant Nos.~ST/P003885/1 and~ST/V001035/1 and
by a Leverhulme Trust Research Project Grant. B.C.\ Backes was supported by STFC through her PhD studentship under grant ST/W50791X/1.
W.N.\ was supperted by the  U.S. Department of Energy, Office of Science, Office of Nuclear Physics under grants DE-SC0013365 and DE-SC0021176 (Office of Advanced Scientific Computing Research and Office of Nuclear Physics, Scientific Discovery through Advanced Computing).
We acknowledge the CSC-IT Center for Science
Ltd., Finland, and the IFT Computer Center of the University of Warsaw, Poland,
for the allocation of computational resources.
This project was partly undertaken on the Viking Cluster,
which is a high-performance computing facility provided by the
University of York. We are grateful for computational support
from the University of York High Performance Computing
service, Viking and the Research Computing team.
We thank Grammarly\textsuperscript{\textregistered} for its support with English writing.
\end{acknowledgments}

\vspace{0.1cm}
\nocite{database}
\bibliography{refs,jacwit42}

@article{frauendorf2014overview,
  title={Overview of neutron--proton pairing},
  author={Frauendorf, Stefan and Macchiavelli, Augusto O},
  journal={Progress in Particle and Nuclear Physics},
  volume={78},
  pages={24--90},
  year={2014},
  publisher={Elsevier}
}

@article{SATULA19971,
title = "Competition between T = 0 and T = 1 pairing in proton-rich nuclei",
journal = "Phys. Lett. B",
volume = "393",
number = "1",
pages = "1 - 6",
year = "1997",
issn = "0370-2693",
doi = "10.1016/S0370-2693(96)01603-6",
url = "http://www.sciencedirect.com/science/article/pii/S0370269396016036",
author = "W. Satu\l{}a and R. Wyss"
}

@article{PRITYCHENKO20161,
title = {Tables of E2 transition probabilities from the first 2+ states in even–even nuclei},
journal = {Atomic Data and Nuclear Data Tables},
volume = {107},
pages = {1-139},
year = {2016},
issn = {0092-640X},
doi = {https://doi.org/10.1016/j.adt.2015.10.001},
url = {https://www.sciencedirect.com/science/article/pii/S0092640X15000406},
author = {B. Pritychenko and M. Birch and B. Singh and M. Horoi}
}

@article{PhysRevLett.117.062501,
  title = {Superdeformed and Triaxial States in $^{42}\mathrm{Ca}$},
  author = {Hady\ifmmode \acute{n}\else \'{n}\fi{}ska-Kl\ifmmode \mbox{\c{e}}\else \c{e}\fi{}k, K. and Napiorkowski, P. J. and Zieli\ifmmode \acute{n}\else \'{n}\fi{}ska, M. and Srebrny, J. and Maj, A. and Azaiez, F. and Valiente Dob\'on, J. J. and Kici\ifmmode \acute{n}\else \'{n}\fi{}ska-Habior, M. and Nowacki, F. and Na\"{\i}dja, H. and Bounthong, B. and Rodr\'{\i}guez, T. R. and de Angelis, G. and Abraham, T. and Anil Kumar, G. and Bazzacco, D. and Bellato, M. and Bortolato, D. and Bednarczyk, P. and Benzoni, G. and Berti, L. and Birkenbach, B. and Bruyneel, B. and Brambilla, S. and Camera, F. and Chavas, J. and Cederwall, B. and Charles, L. and Ciema\l{}a, M. and Cocconi, P. and Coleman-Smith, P. and Colombo, A. and Corsi, A. and Crespi, F. C. L. and Cullen, D. M. and Czermak, A. and D\'esesquelles, P. and Doherty, D. T. and Dulny, B. and Eberth, J. and Farnea, E. and Fornal, B. and Franchoo, S. and Gadea, A. and Giaz, A. and Gottardo, A. and Grave, X. and Gr\ifmmode \mbox{\c{e}}\else \c{e}\fi{}bosz, J. and G\"orgen, A. and Gulmini, M. and Habermann, T. and Hess, H. and Isocrate, R. and Iwanicki, J. and Jaworski, G. and Judson, D. S. and Jungclaus, A. and Karkour, N. and Kmiecik, M. and Karpi\ifmmode \acute{n}\else \'{n}\fi{}ski, D. and Kisieli\ifmmode \acute{n}\else \'{n}\fi{}ski, M. and Kondratyev, N. and Korichi, A. and Komorowska, M. and Kowalczyk, M. and Korten, W. and Krzysiek, M. and Lehaut, G. and Leoni, S. and Ljungvall, J. and Lopez-Martens, A. and Lunardi, S. and Maron, G. and Mazurek, K. and Menegazzo, R. and Mengoni, D. and Merch\'an, E. and M\ifmmode \mbox{\c{e}}\else \c{e}\fi{}czy\ifmmode \acute{n}\else \'{n}\fi{}ski, W. and Michelagnoli, C. and Mierzejewski, J. and Million, B. and Myalski, S. and Napoli, D. R. and Nicolini, R. and Niikura, M. and Obertelli, A. and \"Ozmen, S. F. and Palacz, M. and Pr\'ochniak, L. and Pullia, A. and Quintana, B. and Rampazzo, G. and Recchia, F. and Redon, N. and Reiter, P. and Rosso, D. and Rusek, K. and Sahin, E. and Salsac, M.-D. and S\"oderstr\"om, P.-A. and Stefan, I. and St\'ezowski, O. and Stycze\ifmmode \acute{n}\else \'{n}\fi{}, J. and Theisen, Ch. and Toniolo, N. and Ur, C. A. and Vandone, V. and Wadsworth, R. and Wasilewska, B. and Wiens, A. and Wood, J. L. and Wrzosek-Lipska, K. and Zi\ifmmode \mbox{\c{e}}\else \c{e}\fi{}bli\ifmmode \acute{n}\else \'{n}\fi{}ski, M.},
  journal = {Phys. Rev. Lett.},
  volume = {117},
  issue = {6},
  pages = {062501},
  numpages = {7},
  year = {2016},
  month = {Aug},
  publisher = {American Physical Society},
  doi = {10.1103/PhysRevLett.117.062501},
  url = {https://link.aps.org/doi/10.1103/PhysRevLett.117.062501}
}

@article{PhysRevC.99.044317,
  title = {Probing isospin symmetry in the $(^{50}\mathrm{Fe}, ^{50}\mathrm{Mn}, ^{50}\mathrm{Cr})$ isobaric triplet via electromagnetic transition rates},
  author = {Giles, M. M. and Nara Singh, B. S. and Barber, L. and Cullen, D. M. and Mallaburn, M. J. and Beckers, M. and Blazhev, A. and Braunroth, T. and Dewald, A. and Fransen, C. and Goldkuhle, A. and Jolie, J. and Mammes, F. and M\"uller-Gatermann, C. and W\"olk, D. and Zell, K. O. and Lenzi, S. M. and Poves, A.},
  journal = {Phys. Rev. C},
  volume = {99},
  issue = {4},
  pages = {044317},
  numpages = {7},
  year = {2019},
  month = {Apr},
  publisher = {American Physical Society},
  doi = {10.1103/PhysRevC.99.044317},
  url = {https://link.aps.org/doi/10.1103/PhysRevC.99.044317}
}

@article{PhysRevC.104.029901,
  title = {Erratum: Probing isospin symmetry in the $(^{50}\mathrm{Fe},\phantom{\rule{0.28em}{0ex}}^{50}\mathrm{Mn},\phantom{\rule{0.28em}{0ex}}^{50}\mathrm{Cr})$ isobaric triplet via electromagnetic transition rates [Phys. Rev. C 99, 044317 (2019)]},
  author = {Giles, M. M. and Nara Singh, B. S. and Barber, L. and Cullen, D. M. and Mallaburn, M. J. and Beckers, M. and Blazhev, A. and Braunroth, T. and Dewald, A. and Fransen, C. and Goldkuhle, A. and Jolie, J. and Mammes, F. and M\"uller-Gatermann, C. and W\"olk, D. and Zell, K. O. and Lenzi, S. M. and Poves, A.},
  journal = {Phys. Rev. C},
  volume = {104},
  issue = {2},
  pages = {029901},
  numpages = {1},
  year = {2021},
  month = {Aug},
  publisher = {American Physical Society},
  doi = {10.1103/PhysRevC.104.029901},
  url = {https://link.aps.org/doi/10.1103/PhysRevC.104.029901}
}

@article{WIMMER2023138249,
title = {Isospin symmetry in the {$T=1,A=62$} triplet},
journal = {Physics Letters B},
volume = {847},
pages = {138249},
year = {2023},
issn = {0370-2693},
doi = {https://doi.org/10.1016/j.physletb.2023.138249},
url = {https://www.sciencedirect.com/science/article/pii/S037026932300583X},
author = {K. Wimmer and P. Ruotsalainen and S.M. Lenzi and A. Poves and T. Hüyük and F. Browne and P. Doornenbal and T. Koiwai and T. Arici and K. Auranen and M.A. Bentley and M.L. Cortés and C. Delafosse and T. Eronen and Z. Ge and T. Grahn and P.T. Greenlees and A. Illana and N. Imai and H. Joukainen and R. Julin and A. Jungclaus and H. Jutila and A. Kankainen and N. Kitamura and B. Longfellow and J. Louko and R. Lozeva and M. Luoma and B. Mauss and D.R. Napoli and M. Niikura and J. Ojala and J. Pakarinen and X. Pereira-Lopez and P. Rahkila and F. Recchia and M. Sandzelius and J. Sarén and R. Taniuchi and H. Tann and S. Uthayakumaar and J. Uusitalo and V. Vaquero and R. Wadsworth and G. Zimba and R. Yajzey},
}

@article{MORSE2018198,
title = {Lifetime measurement of the {21$^+$} state in {$^{74}$Rb} and isospin properties of quadrupole transition strengths at $N=Z$},
journal = {Physics Letters B},
volume = {787},
pages = {198-203},
year = {2018},
issn = {0370-2693},
doi = {https://doi.org/10.1016/j.physletb.2018.10.064},
url = {https://www.sciencedirect.com/science/article/pii/S037026931830844X},
author = {C. Morse and H. Iwasaki and A. Lemasson and A. Dewald and T. Braunroth and V.M. Bader and T. Baugher and D. Bazin and J.S. Berryman and C.M. Campbell and A. Gade and C. Langer and I.Y. Lee and C. Loelius and E. Lunderberg and F. Recchia and D. Smalley and S.R. Stroberg and R. Wadsworth and C. Walz and D. Weisshaar and A. Westerberg and K. Whitmore and K. Wimmer}
}

@article{Romero,
title = "Symmetry restoration in the mean-field description of proton-neutron pairing",
journal = "Phys. Lett. B",
volume = "795",
pages = "177 - 182",
year = "2019",
issn = "0370-2693",
doi = "10.1016/j.physletb.2019.06.032",
url = "http://www.sciencedirect.com/science/article/pii/S0370269319304113",
author = "Romero, A. M. and J. Dobaczewski and A. Pastore"
}

@article{martinez,
title = "Competition of isoscalar and isovector proton-neutron pairing in nuclei",
journal = "Nucl. Phys. A",
volume = "651",
number = "4",
pages = "379 - 393",
year = "1999",
issn = "0375-9474",
doi = "10.1016/S0375-9474(99)00141-4",
url = "http://www.sciencedirect.com/science/article/pii/S0375947499001414",
author = "G. Mart\'{i}nez-Pinedo and K. Langanke and P. Vogel"
}

@article{Engel1997,
  author = {Engel, J. and Pittel, S. and Stoitsov, M. and Vogel, P. and Dukelsky, J.},
  journal = {Phys. Rev. C},
  volume = {55},
  issue = {4},
  pages = {1781--1788},
  numpages = {0},
  year = {1997},
  publisher = {American Physical Society},
  doi = {10.1103/PhysRevC.55.1781},
  url = {https://link.aps.org/doi/10.1103/PhysRevC.55.1781}
}

@article{Wimmer21,
  title = {Shape Changes in the Mirror Nuclei $^{70}\mathrm{Kr}$ and $^{70}\mathrm{Se}$},
  author = {Wimmer, K. and Korten, W. and Doornenbal, P. and Arici, T. and Aguilera, P. and Algora, A. and Ando, T. and Baba, H. and Blank, B. and Boso, A. and Chen, S. and Corsi, A. and Davies, P. and de Angelis, G. and de France, G. and Delaroche, J.-P. and Doherty, D. T. and Gerl, J. and Gernh\"auser, R. and Girod, M. and Jenkins, D. and Koyama, S. and Motobayashi, T. and Nagamine, S. and Niikura, M. and Obertelli, A. and Libert, J. and Lubos, D. and Rodr\'{\i}guez, T. R. and Rubio, B. and Sahin, E. and Saito, T. Y. and Sakurai, H. and Sinclair, L. and Steppenbeck, D. and Taniuchi, R. and Wadsworth, R. and Zielinska, M.},
  journal = {Phys. Rev. Lett.},
  volume = {126},
  issue = {7},
  pages = {072501},
  numpages = {6},
  year = {2021},
  month = {Feb},
  publisher = {American Physical Society},
  doi = {10.1103/PhysRevLett.126.072501},
  url = {https://link.aps.org/doi/10.1103/PhysRevLett.126.072501}
}

@article{Boso,
title = {Isospin dependence of electromagnetic transition strengths among an isobaric triplet},
journal = {Physics Letters B},
volume = {797},
pages = {134835},
year = {2019},
issn = {0370-2693},
doi = {https://doi.org/10.1016/j.physletb.2019.134835},
url = {https://www.sciencedirect.com/science/article/pii/S0370269319305490},
author = {A. Boso and S.A. Milne and M.A. Bentley and F. Recchia and S.M. Lenzi and D. Rudolph and M. Labiche and X. Pereira-Lopez and S. Afara and F. Ameil and T. Arici and S. Aydin and M. Axiotis and D. Barrientos and G. Benzoni and B. Birkenbach and A.J. Boston and H.C. Boston and P. Boutachkov and A. Bracco and A.M. Bruce and B. Bruyneel and B. Cederwall and E. Clement and M.L. Cortes and D.M. Cullen and P. Désesquelles and Zs. Dombràdi and C. Domingo-Pardo and J. Eberth and C. Fahlander and M. Gelain and V. González and P.R. John and J. Gerl and P. Golubev and M. Górska and A. Gottardo and T. Grahn and L. Grassi and T. Habermann and L.J. Harkness-Brennan and T.W. Henry and H. Hess and I. Kojouharov and W. Korten and N. Lalović and M. Lettmann and C. Lizarazo and C. Louchart-Henning and R. Menegazzo and D. Mengoni and E. Merchan and C. Michelagnoli and B. Million and V. Modamio and T. Moeller and D.R. Napoli and J. Nyberg and B.S. {Nara Singh} and H. Pai and N. Pietralla and S. Pietri and Zs. Podolyak and R.M. {Perez Vidal} and A. Pullia and D. Ralet and G. Rainovski and M. Reese and P. Reiter and M.D. Salsac and E. Sanchis and L.G. Sarmiento and H. Schaffner and L.M. Scruton and P.P. Singh and C. Stahl and S. Uthayakumaar and J.J. Valiente-Dobón and O. Wieland}
}

@article{Giles,
  title = {Probing isospin symmetry in the $(^{50}\mathrm{Fe}, ^{50}\mathrm{Mn}, ^{50}\mathrm{Cr})$ isobaric triplet via electromagnetic transition rates},
  author = {Giles, M. M. and Nara Singh, B. S. and Barber, L. and Cullen, D. M. and Mallaburn, M. J. and Beckers, M. and Blazhev, A. and Braunroth, T. and Dewald, A. and Fransen, C. and Goldkuhle, A. and Jolie, J. and Mammes, F. and M\"uller-Gatermann, C. and W\"olk, D. and Zell, K. O. and Lenzi, S. M. and Poves, A.},
  journal = {Phys. Rev. C},
  volume = {99},
  issue = {4},
  pages = {044317},
  numpages = {7},
  year = {2019},
  month = {Apr},
  publisher = {American Physical Society},
  doi = {10.1103/PhysRevC.99.044317},
  url = {https://link.aps.org/doi/10.1103/PhysRevC.99.044317}
}

@article{Bernstein,
  title = {Isospin Decomposition of Nuclear Multipole Matrix Elements from $\ensuremath{\gamma}$ Decay Rates of Mirror Transitions: Test of Values Obtained with Hadronic Probes},
  author = {Bernstein, Aron M. and Brown, V. R. and Madsen, V. A.},
  journal = {Phys. Rev. Lett.},
  volume = {42},
  issue = {7},
  pages = {425--428},
  numpages = {0},
  year = {1979},
  month = {Feb},
  publisher = {American Physical Society},
  doi = {10.1103/PhysRevLett.42.425},
  url = {https://link.aps.org/doi/10.1103/PhysRevLett.42.425}
}

@article{Warb,
journal = {in Isospin in Nuclear Physics edited by D. H. Wilkinson (North-Holland, Amsterdam)},
volume = {Ch. 5},
year = {1969},
author = {E. K. Warburton and J. Weneser},
}

@inproceedings{Muir:2020sft,
    author = "Muir, David and Pr\'ochniak, Leszek and Pastore, Alessandro and Dobaczewski, Jacek",
    title = "{Structure of Krypton Isotopes using the Generalised Bohr Hamiltonian Method}",
    booktitle = "{27th International Nuclear Physics Conference}",
    eprint = "2004.10835",
    archivePrefix = "arXiv",
    primaryClass = "nucl-th",
    doi = "10.1088/1742-6596/1643/1/012147",
    month = "4",
    year = "2020"
}

@article{gir79,
  title={The zero-point energy correction and its effect on nuclear dynamics},
  author={Girod, M and Grammaticos, B},
  journal={Nuclear Physics A},
  volume={330},
  number={1},
  pages={40--52},
  year={1979},
  publisher={Elsevier}
}

@article{bar11,
  title={Quadrupole collective inertia in nuclear fission: Cranking approximation},
  author={Baran, A and Sheikh, JA and Dobaczewski, J and Nazarewicz, Witold and Staszczak, A},
  journal={Physical Review C},
  volume={84},
  number={5},
  pages={054321},
  year={2011},
  publisher={APS}
}

@article{dob09,
  title={Solution of the {Skyrme-Hartree-Fock-Bogolyubov} equations in the {Cartesian} deformed harmonic-oscillator {basis.:(VI) hfodd (v2. 40h): A} new version of the program},
  author={Dobaczewski, Jacek and others},
  journal={Computer Physics Communications},
  volume={180},
  number={11},
  pages={2361--2391},
  year={2009},
  publisher={Elsevier}
}

@book{row10,
  title={Fundamentals of nuclear models: foundational models},
  author={Rowe, David J and Wood, John L},
  year={2010},
  publisher={World Scientific Publishing Company}
}

@article{pro15,
  title={Microscopic description of collective properties of even-even Xe isotopes},
  author={Prochniak, L},
  journal={Physica Scripta},
  volume={90},
  number={11},
  pages={114005},
  year={2015},
  publisher={IOP Publishing}
}

@article{pro09,
  title={Quadrupole collective states within the Bohr collective Hamiltonian},
  author={Pr{\'o}chniak, L and Rohozi{\'n}ski, SG},
  journal={J. Phys. G},
  volume={36},
  number={12},
  pages={123101},
  year={2009},
  publisher={IOP Publishing}
}

@article{Hagino2002,
title={Projection and ground state correlations made simple},
author={K. Hagino and P.--G. Reinhard and G.F. Bertsch},
journal={Phys. Rev. C },
volume={65},
year={2002},
pages={064320},
note={http://www.arxiv.org/abs/nucl-th/0202079},
url={http://link.aps.org/doi/10.1103/PhysRevC.65.064320},
}

@article{Reinhard1978,
title={The zero-point energy for rotation},
author={P.--G. Reinhard},
journal={Z. Phys. A },
volume={285},
year={1978},
pages={93 }
}

@article{Reinhard1987,
title={The generator coordinate method and quantised collective motion in     nuclear systems },
author={P.--G. Reinhard  and K. Goeke},
journal={Rep. Prog. Phys. },
volume={50},
year={1987},
pages={1 },
url={http://iopscience.iop.org/0034-4885/50/1/001}
}

@misc{database,
  title = {Database for the findings of this article},
  howpublished = {\url{https://webfiles.york.ac.uk/HFODD/Projects/Isobaric_Triplets/}}
}

@article{Reinhard1999,
  title = {Shape coexistence and the effective nucleon-nucleon interaction},
  author = {Reinhard, P.-G. and Dean, D. J. and Nazarewicz, W. and Dobaczewski, J. and Maruhn, J. A. and Strayer, M. R.},
  journal = {Phys. Rev. C},
  volume = {60},
  issue = {1},
  pages = {014316},
  numpages = {20},
  year = {1999},
  month = {Jun},
  publisher = {American Physical Society},
  doi = {10.1103/PhysRevC.60.014316},
  url = {https://link.aps.org/doi/10.1103/PhysRevC.60.014316}
}

@Article{Goz85a,
  author   = "A. Gozdz and K. Pomorski and M. Brack and W. Werner",
  title    = "The Mass Parameters for the Average Mean-Field Potential",
  journal  = {Nucl. Phys. A},
  volume   = "442",
  year     = "1985",
  pages    = "26",
}

@article{Erler2011,
  title ={Self-consistent nuclear mean-field models:
          example Skyrme-Hartree-Fock},
  author =  {J. Erler and P. Kl\"upfel and P.-G. Reinhard},
  journal = {J. Phys. G},
  volume =  {38},                                 
  year =    {2011},
  pages =   {033101},
  url={http://dx.doi.org/10.1088/0954-3899/38/3/033101},
}

@article{Reinhard2021,
  title =   {The Axial Hartree–Fock + BCS Code SkyAx},
  author =  {P.-G. Reinhard and B. Schuetrumpf and J.A. Maruhn},
  journal = {Comp. Phys. Comm.},
  volume =  {258},
  year =    {2021},
  pages =   {107603},
  url={https://doi.org/10.1016/j.cpc.2020.107603}
}

@article{Kluepfel2008,
title={Systematics of collective correlation energies from self-consistent mean-field calculations},
author={P. Kl\"upfel and J. Erler and P.--G. Reinhard and J. A. Maruhn},
journal={Eur. Phys. J A},
volume={37},
year={2008},
pages={343},
note={http://www.arxiv.org/abs/0804.340},
url={http://dx.doi.org/10.1140/epja/i2008-10633-3},
}

@Article{Kri90a,
  author =       "S. J. Krieger and P. Bonche and H. Flocard and
                  P. Quentin and M. S. Weiss",
 title =         "An Improved Pairing Interaction for Mean--Field
                  Calculations using Skyrme Potentials",
 journal =       {Nucl. Phys. A},
 volume =        "517",
 year =          1990,
 pages =         "275"
}

@Inbook{Lenzi2009,
author="Lenzi, S.M.
and Bentley, M.A.",
editor="Al-Khalili, J.S.
and Roeckl, E.",
title="Test of Isospin Symmetryisospin symmetry Along the {$N=Z$} Line",
bookTitle="The Euroschool Lectures on Physics with Exotic Beams, Vol. III",
year="2009",
publisher="Springer Berlin Heidelberg",
address="Berlin, Heidelberg",
pages="57--98",
isbn="978-3-540-85839-3",
doi="10.1007/978-3-540-85839-3_3",
url="https://doi.org/10.1007/978-3-540-85839-3_3"
}

@Article{Smirnova2023,
AUTHOR = {Smirnova, Nadezda A.},
TITLE = {Isospin-Symmetry Breaking within the Nuclear Shell {Model: Present }Status and Developments},
JOURNAL = {Physics},
VOLUME = {5},
YEAR = {2023},
NUMBER = {2},
PAGES = {352--380},
URL = {https://www.mdpi.com/2624-8174/5/2/26},
ISSN = {2624-8174},
DOI = {10.3390/physics5020026}
}

@Article{Bentley2022,
AUTHOR = {Bentley, Michael A},
TITLE = {Excited States in Isobaric Multiplets—Experimental Advances and the Shell-Model Approach},
JOURNAL = {Physics},
VOLUME = {4},
YEAR = {2022},
NUMBER = {3},
PAGES = {995--1011},
URL = {https://www.mdpi.com/2624-8174/4/3/66},
DOI = {10.3390/physics4030066}
}

@article{Satula2001,
  title = {Microscopic Structure of Fundamental Excitations in $\mathit{N}\phantom{\rule{0ex}{0ex}}=\phantom{\rule{0ex}{0ex}}\mathit{Z}$ Nuclei},
  author = {Satu\l{}a, Wojciech and Wyss, Ramon},
  journal = {Phys. Rev. Lett.},
  volume = {87},
  issue = {5},
  pages = {052504},
  numpages = {4},
  year = {2001},
  month = {Jul},
  publisher = {American Physical Society},
  doi = {10.1103/PhysRevLett.87.052504},
  url = {https://link.aps.org/doi/10.1103/PhysRevLett.87.052504}
}

@article{Glowacz2004,
	author = {G{\l}owacz, S. and Satu{\l}a, W. and Wyss, R. A.},
	doi = {10.1140/epja/i2003-10111-6},
	journal = {Eur. Phys. J. A},
	number = {1},
	pages = {33--44},
	title = {Cranking in isospace},
	url = {https://doi.org/10.1140/epja/i2003-10111-6},
	volume = {19},
	year = {2004},
}

@article{Cwiok1980,
	author = {S. \'Cwiok and  W. Nazarewicz and  W. Zych},
	journal = {Acta. Phys. Pol. B},
	pages = {445},
	title = {The Dependence of {Coulomb }Displacement Energy on Deformation of a Nucleus},
	url = {https://www.actaphys.uj.edu.pl/fulltext?series=Reg&vol=11&page=445},
	volume = {11},
	year = {1980},
}

@article{Sheikh_2021,
doi = {10.1088/1361-6471/ac288a},
url = {https://dx.doi.org/10.1088/1361-6471/ac288a},
year = {2021},
month = {nov},
publisher = {IOP Publishing},
volume = {48},
number = {12},
pages = {123001},
author = {J A Sheikh and J Dobaczewski and P Ring and L M Robledo and C Yannouleas},
title = {Symmetry restoration in mean-field approaches},
journal = {J. Phys. G},
}

@article{Nazarewicz1985,
title = "Microscopic study of the high-spin behaviour in selected ${A} \simeq 80$ nuclei",
journal = "Nucl. Phys. A",
volume = "435",
number = "2",
pages = "397 - 447",
year = "1985",
issn = "0375-9474",
doi = "10.1016/0375-9474(85)90471-3",
url = "http://www.sciencedirect.com/science/article/pii/0375947485904713",
author = "W. Nazarewicz and J. Dudek and R. Bengtsson and T. Bengtsson and I. Ragnarsson",
}

@Proceedings{ZirconiumRegion1988,
editor="Eberth, J{\"u}rgen
and Meyer, Richard A.
and Sistemich, Kornelius",
title="Nuclear Structure of the Zirconium Region",
year="1988",
publisher="Springer Berlin Heidelberg",
address="Berlin, Heidelberg",
isbn="978-3-642-73958-3",
url={https://link.springer.com/book/10.1007/978-3-642-73958-3}
}

@article{Satula2012,
  title = {Isospin-breaking corrections to superallowed Fermi $\ensuremath{\beta}$ decay in isospin- and angular-momentum-projected nuclear density functional theory},
  author = {Satu\l{}a, W. and Dobaczewski, J. and Nazarewicz, W. and Werner, T. R.},
  journal = {Phys. Rev. C},
  volume = {86},
  issue = {5},
  pages = {054316},
  numpages = {13},
  year = {2012},
  month = {Nov},
  publisher = {American Physical Society},
  doi = {10.1103/PhysRevC.86.054316},
  url = {https://link.aps.org/doi/10.1103/PhysRevC.86.054316}
}

@article{Satula2011,
  title = {Microscopic Calculations of Isospin-Breaking Corrections to Superallowed Beta Decay},
  author = {Satu\l{}a, W. and Dobaczewski, J. and Nazarewicz, W. and Rafalski, M.},
  journal = {Phys. Rev. Lett.},
  volume = {106},
  issue = {13},
  pages = {132502},
  numpages = {4},
  year = {2011},
  month = {Mar},
  publisher = {American Physical Society},
  doi = {10.1103/PhysRevLett.106.132502},
  url = {https://link.aps.org/doi/10.1103/PhysRevLett.106.132502}
}

@article{Bertsch2009,
  title = {Odd-even mass differences from self-consistent mean field theory},
  author = {Bertsch, G. F. and Bertulani, C. A. and Nazarewicz, W. and Schunck, N. and Stoitsov, M. V.},
  journal = {Phys. Rev. C},
  volume = {79},
  issue = {3},
  pages = {034306},
  numpages = {12},
  year = {2009},
  month = {Mar},
  publisher = {American Physical Society},
  doi = {10.1103/PhysRevC.79.034306},
  url = {https://link.aps.org/doi/10.1103/PhysRevC.79.034306}
}

@article{Anguiano2001,
title = {Coulomb exchange and pairing contributions in nuclear{ Hartree-–Fock-–Bogoliubov calculations with the Gogny}  force},
journal = {Nucl. Phys. A},
volume = {683},
number = {1},
pages = {227--254},
year = {2001},
issn = {0375-9474},
doi = {10.1016/S0375-9474(00)00445-0},
url = {https://www.sciencedirect.com/science/article/pii/S0375947400004450},
author = {M. Anguiano and J.L. Egido and L.M. Robledo},
}

@article{Belyaev1959,
title = {Effect of pairing correlations on nuclear properties},
journal = {Mat. Fys. Medd. Dan. Vid. Selsk.},
volume = {31},
number = {11},
year = {1959},
url = {https://gymarkiv.sdu.dk/MFM/kdvs/mfm%2030-39/mfm-31-11.pdf},
author = {S.T. Belyaev},
}

@BOOK{Bohr75,
   author = "Bohr, A. and Mottelson, B. R.",
   title = "Nuclear Structure, vol. II",
   year = {1975},
   publisher ={W. A. Benjamin, Reading}
}

@article{Brack1972,
  title = {Funny Hills: {The} Shell-Correction Approach to Nuclear Shell Effects and Its Applications to the Fission Process},
  author = {Brack, M. and Damgaard, Jens and Jensen, A. S. and Pauli, H. C. and Strutinsky, V. M. and Wong, C. Y.},
  journal = {Rev. Mod. Phys.},
  volume = {44},
  issue = {2},
  pages = {320--405},
  numpages = {0},
  year = {1972},
  month = {Apr},
  publisher = {American Physical Society},
  doi = {10.1103/RevModPhys.44.320},
  url = {https://link.aps.org/doi/10.1103/RevModPhys.44.320}
}

@misc{ENSDF2,
note = "Evaluated Nuclear Structure Data File (ENSDF),
http://www.nndc.bnl.gov/ensdf/"
}

@article{Reinhard2021a,
  title = {Nuclear charge densities in spherical and deformed nuclei: Toward precise calculations of charge radii},
  author = {Reinhard, Paul-Gerhard and Nazarewicz, Witold},
  journal = {Phys. Rev. C},
  volume = {103},
  issue = {5},
  pages = {054310},
  numpages = {9},
  year = {2021},
  month = {May},
  publisher = {American Physical Society},
  doi = {10.1103/PhysRevC.103.054310},
  url = {https://link.aps.org/doi/10.1103/PhysRevC.103.054310}
}

\end{document}